\documentclass[cits]{PoS}

\newcommand{\xref}[2]{\href{http://arxiv.org/abs/#1}{arXiv:{#1} {#2}}}

\newcommand{\oref}[1]{\href{http://arxiv.org/abs/#1}{#1}}

\newcommand{\dref}[2]{\href{http://dx.doi.org/#1}{#2}}

\newcommand{\ra}{\rightarrow}

\title{Non-QCD contributions to top-pair production near threshold\thanks{TUM-HEP-1024/15}}

\ShortTitle{Non-QCD contributions to top pair production near threshold}

\author{Martin Beneke\\
Physik Department T31, James-Franck-Stra{\ss}e 1, 
Technische Universit\"at M\"unchen,\\ 85748 Garching, Germany\\
}
        
\author{Andreas Maier\\
IPPP, Department of Physics, University of Durham, DH1 3LE, United Kingdom\\
}
        
\author{Jan Piclum\\
Albert Einstein Center for Fundamental Physics, 
Institute for Theoretical Physics,\\ Sidlerstrasse 5, CH-3012 Bern, 
Switzerland\\
}

\author{\speaker{Thomas Rauh}\\
Physik Department T31, James-Franck-Stra{\ss}e 1, 
Technische Universit\"at M\"unchen,\\ 85748 Garching, Germany\\
E-mail: \email{thomas.rauh@mytum.de}}

\abstract{\vskip0.4cm
The threshold scan of top pair production at a future lepton collider allows 
to determine several Standard Model parameters with very high precision. 
The recent completion of the third-order QCD corrections to the inclusive 
top-pair production cross section demonstrated that strong dynamics are under 
control. We investigate effects from P-wave production and Higgs contributions 
at third order and from QED and the nonresonant production of the physical 
$W^+W^-b\bar{b}$ final state at first order. We discuss the sensitivity of 
the cross section to the top mass, width and Yukawa coupling as well as to 
the strong coupling.}

\FullConference{The European Physical Society Conference on High Energy 
Physics\\
22-29 July 2015\\
Vienna, Austria}

\begin{document}

\section{Introduction}
%\textit{Introduction.}
\noindent The top-quark mass is an important parameter for many
observables in the Standard Model and beyond, including the stability
of the vacuum, due to the often large radiative 
corrections involving virtual top quarks. 
%This makes knowledge of its precise value mandatory for tests
%of the former and searches for the latter. 
Currently, the highest
precision is achieved in the direct reconstruction of (anti-) top
quarks from their decay products at the Tevatron and LHC with a total
uncertainty of $\pm 0.76$~GeV~\cite{ATLAS:2014wva}. However, this
value is plagued by the lack of understanding of the precise
relationship between the measured Monte Carlo mass and a ``proper''
mass definition from the theory point of view, which could add an
additional uncertainty of the order of 1~GeV. This can be circumvented
by determinations of the top-quark mass from the measurement of 
the total top pair production cross section in hadron
collisions~\cite{Aad:2014kva,Chatrchyan:2013haa} or indirectly, from 
flavour and electroweak precision observables~\cite{Giudice:2015toa}. The
drawback is an increased uncertainty at the level of several GeV. A
measurement of the top quark mass with an uncertainty substantially below $\pm
1$~GeV can be achieved by performing a threshold scan at a
future lepton collider, which consists of the measurement of the total
inclusive top pair production cross section for about ten
center-of-mass energies near the production threshold
$\sqrt{s}\sim2m_t$~\cite{Martinez:2002st,Seidel:2013sqa,Horiguchi:2013wra}. By
comparison of the shape of the cross section with the theory
prediction the top mass can be measured directly in a well-defined
short-distance mass scheme and with very high accuracy. Furthermore
the top width can be determined precisely and modifications of the 
top Yukawa coupling through new physics effects could possibly be detected. 
For this program it is
crucial that the level of accuracy provided by a lepton collider is
matched on the theory side. The recent completion of the QCD
contributions up to NNNLO precision~\cite{Beneke:2015kwa} showed that
the theory uncertainty is now greatly reduced with respect to the
NNLO predictions~\cite{Hoang:2000yr} and at the level of
just~$\pm3\%$. Thus non-QCD effects, which can affect the cross
section by up to~$10\%$~\cite{Beneke:2015lwa}, have now become the
focus of further theoretical efforts. In the following we give
a very brief outline of the special dynamics near the production
threshold, then discuss various non-QCD effects and present numerical
results for the cross section and its sensitivity to different input
parameters.

\textit{Threshold dynamics.} 
In the vicinity of the top-pair production
threshold $\sqrt{s}\sim2m_t$ the tops are nonrelativistic with a small
velocity of the order of the strong coupling constant $v\sim\alpha_s$.
Thus the top mass $m_t$, momentum $m_tv$ and energy $m_tv^2$ are vastly
different and set the relevant scales, denoted as the hard, soft and 
ultrasoft scale. In addition terms scaling like $(\alpha_s/v)^n$
appear which are not suppressed in the nonrelativistic counting and
indicate the breakdown of conventional perturbation theory. Hence
these so-called Coulomb singularities have to be resummed to all orders.
This can be achieved in the effective field theory of potential 
non-relativistic QCD (PNRQCD), which is obtained by subsequently integrating
out the hard and soft scale. A distinguishing aspect of PNRQCD is that
the LO Lagrangian does not describe free fields, but nonrelativistic top 
fields which are interacting through an instantaneous colour Coulomb potential.
Consequently the leading order Coulomb interaction is treated 
nonperturbatively while higher order corrections can be obtained 
systematically by expanding in $\alpha_s$ and $v$ around the resummed
solution. For more details on the EFT framework we refer to the 
literature~\cite{Beneke:2013jia}, where everything required for the
NNNLO cross section is described.
The cross section, normalized as usual to the muon pair production 
cross section, can be expressed using the optical theorem as
\begin{equation}
 R(s) \equiv \frac{\sigma(e^+e^-\ra \gamma^*, \,Z^* 
\ra t\bar{t}X)}{\sigma_0(e^+e^-\ra\mu^+\mu^-)} = 12\pi\, f(s)
\mbox{ Im}\left[\Pi^{(v)}(s)\right],
\end{equation}
where $f(s) = e_t^2 + \dots$ is a prefactor depending on the top
couplings to photons and $Z$ bosons and kinematic variables.
The vector polarization function $\Pi^{(v)}(s)$ has the form
\begin{equation}
 \Pi^{(v)}(s) = \frac{3}{2m_t^2}c_v\left[c_v-\frac{E}{m_t}
\left(c_v+\frac{d_v}{3}\right)\right]G(E)+\dots,
\end{equation}
where $E=\sqrt{s}-2m_t$ is the energy of the top pair, $c_v,d_v$ are 
hard matching coefficients for the external vector current,
%\footnote{
%Since PNRQCD is a nonrelativistic theory the top and anti-top 
%particle numbers are separately conserved and production/annihilation 
%of top pairs can only be described through external operators.} 
and the Green function $G(E)$ describes the propagation of the top 
pair within PNRQCD, subject to interactions from various 
potentials and the exchange of ultrasoft gluons. The imaginary part 
of the vector polarization function is known to third order in 
the reorganized and resummed expansion in $\alpha_s$ and $v$, see 
Figure~1 of \cite{Beneke:2015kwa}. 

\section{Non-QCD and P-wave contribution}
\noindent In the following we discuss further 
effects not contained in the QCD vector polarization function, which 
are parametrically or numerically of similar size as the remaining $\pm 3\%$ 
theoretical uncertainty on the contribution from $\Pi^{(v)}(s)$.

\textit{P-wave contribution.} 
In addition to the dominant contribution from the vector current 
as described above, the top pair can also be produced through an 
axial-vector current from the exchange of a $s$-channel Z boson. 
This yields top pairs in a P-wave state which are suppressed by 
a factor $v^2$ with respect to the leading S-wave production 
and thus constitutes a NNLO effect. The full contribution up 
to NNNLO has been computed and discussed in~\cite{Beneke:2013kia}.
This correction is only of the order of~$1\%$ relative to the 
third-order S-wave QCD result~\cite{Beneke:2015kwa}, and is included 
in what is referred to as the QCD prediction below.

\textit{Higgs effects.} 
In the following we consider only Higgs effects
that come from the top Yukawa coupling. Contributions involving the
coupling to gauge bosons will be regarded as general electroweak
effects and treated separately. The former manifest themselves 
as corrections to 
the hard matching coefficient $c_v$ of the external vector current and
in a local contribution to the $t\bar{t}$ potential.\footnote{
Earlier work included the potential induced by Higgs exchange 
in the form of a Yukawa potential 
\cite{Strassler:1990nw,Harlander:1995dp}. Consistency with the 
implementation of the matching coefficients requires that the 
Yukawa potential is approximated by a local potential when the Higgs mass 
is much larger than the typical potential momentum exchange, and 
treated as a perturbation.} The pure Higgs contribution to $c_v$ has been 
computed in~\cite{Grzadkowski:1986pm,Guth:1991ab,Eiras:2006xm}
and mixed Higgs and QCD corrections in~\cite{Eiras:2006xm}.
The insertion of the Higgs potential into the Green function
was calculated recently in~\cite{Beneke:2015lwa}, such that the 
full NNNLO Higgs correction to the cross section is now known.

\begin{figure}[t]
  \centering
  \includegraphics[width=0.35\textwidth]{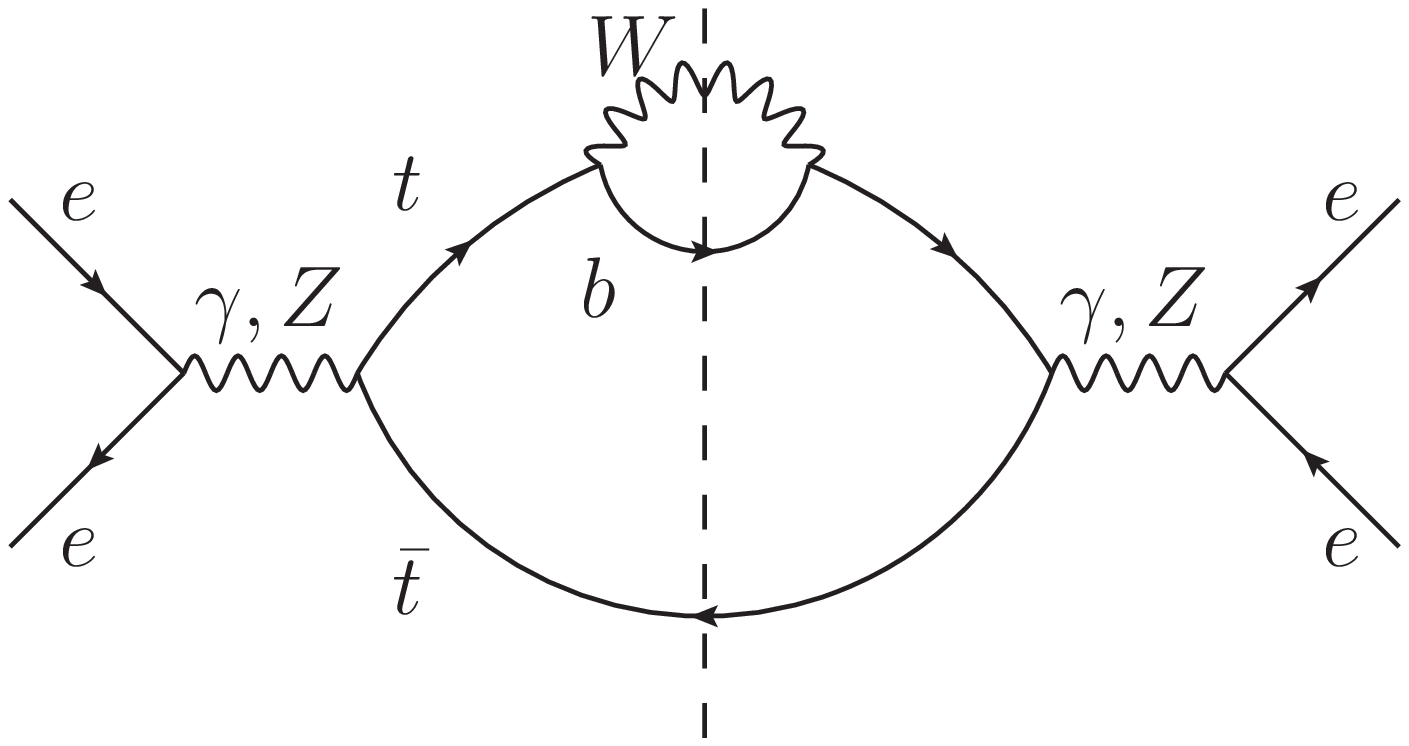} \hspace{0.4cm}
  \includegraphics[width=0.35\textwidth]{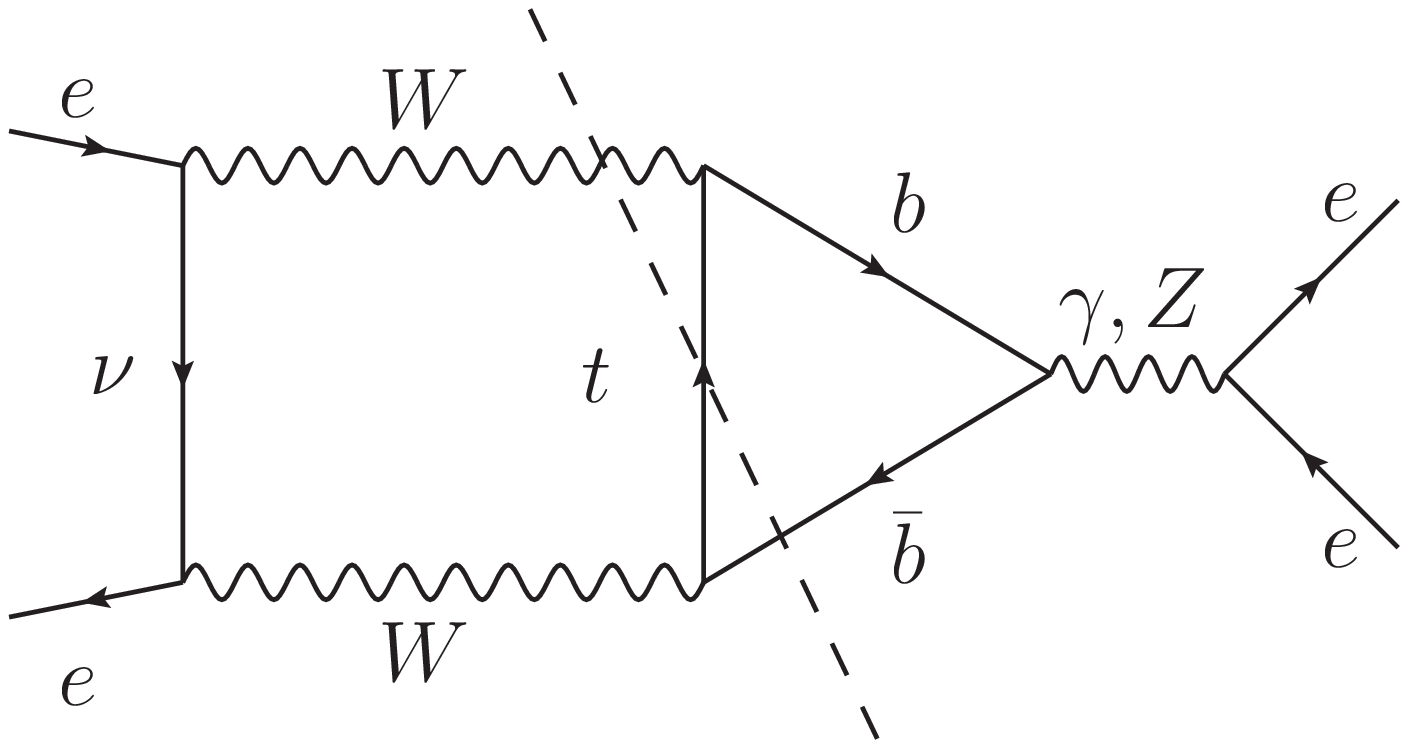}
\caption{\label{fig:nonresonant} Sample diagrams for the nonresonant
production of the $W^+W^-b\bar{b}$ final state.}
\end{figure}

\textit{Nonresonant effects.} Since the cross section near threshold 
is also sensitive to the small ultrasoft scale and the top width is 
of the same order, the narrow-width approximation cannot be used here 
to factorize the production and decay of the top pair. This implies 
that, assuming $V_{tb}=1$, the physical final state is $W^+W^-b\bar{b}$.
It is dominantly produced through a resonant top pair, where the 
replacement $E\rightarrow E+i\Gamma_t$ accounts for the effects of top 
instability~\cite{Fadin:1987wz}. At higher orders in the nonrelativistic 
counting the  $W^+W^-b\bar{b}$ final state can however also be produced 
with just one or no resonant tops. Two sample diagrams at NLO without 
an on-shell top (left) or anti-top (right) are shown in 
Figure~\ref{fig:nonresonant}.
Only the sum of both processes constitutes a physical quantity as is 
also apparent from singularities that appear in both parts at NNLO
and only cancel in the sum. In a systematic way the two contributions 
can be organized within Unstable Particle Effective 
Theory~\cite{Beneke:2003xh,Beneke:2004km}.
The nonresonant NLO effects have been calculated 
in\cite{Beneke:2010mp} and have been included in~\cite{Beneke:2015lwa}.
At NNLO only partial results are available~\cite{Penin:2011gg,Jantzen:2013gpa,
Ruiz-Femenia:2014ava}, which we have not considered yet.

\begin{figure}[t]
  \centering
  \includegraphics[width=0.48\textwidth]{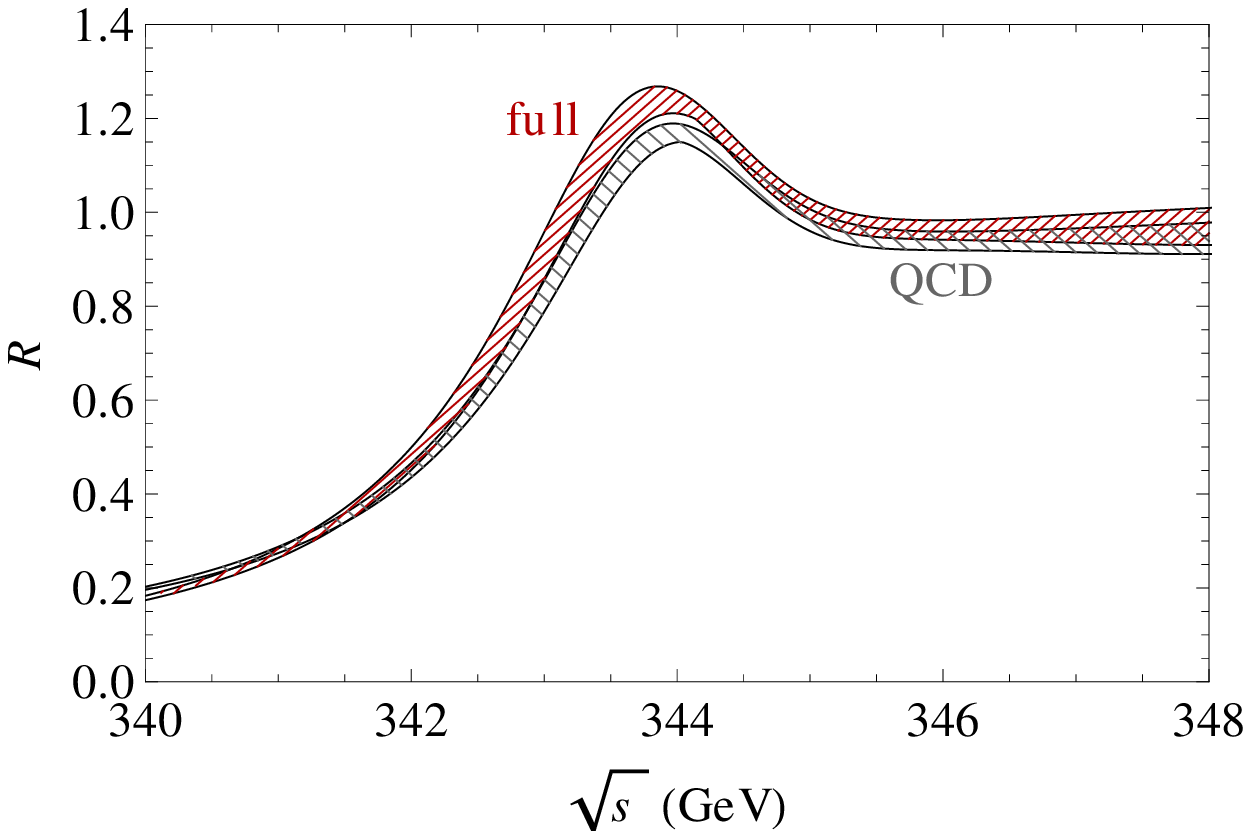}\hspace{0.2cm}
  \includegraphics[width=0.48\textwidth]{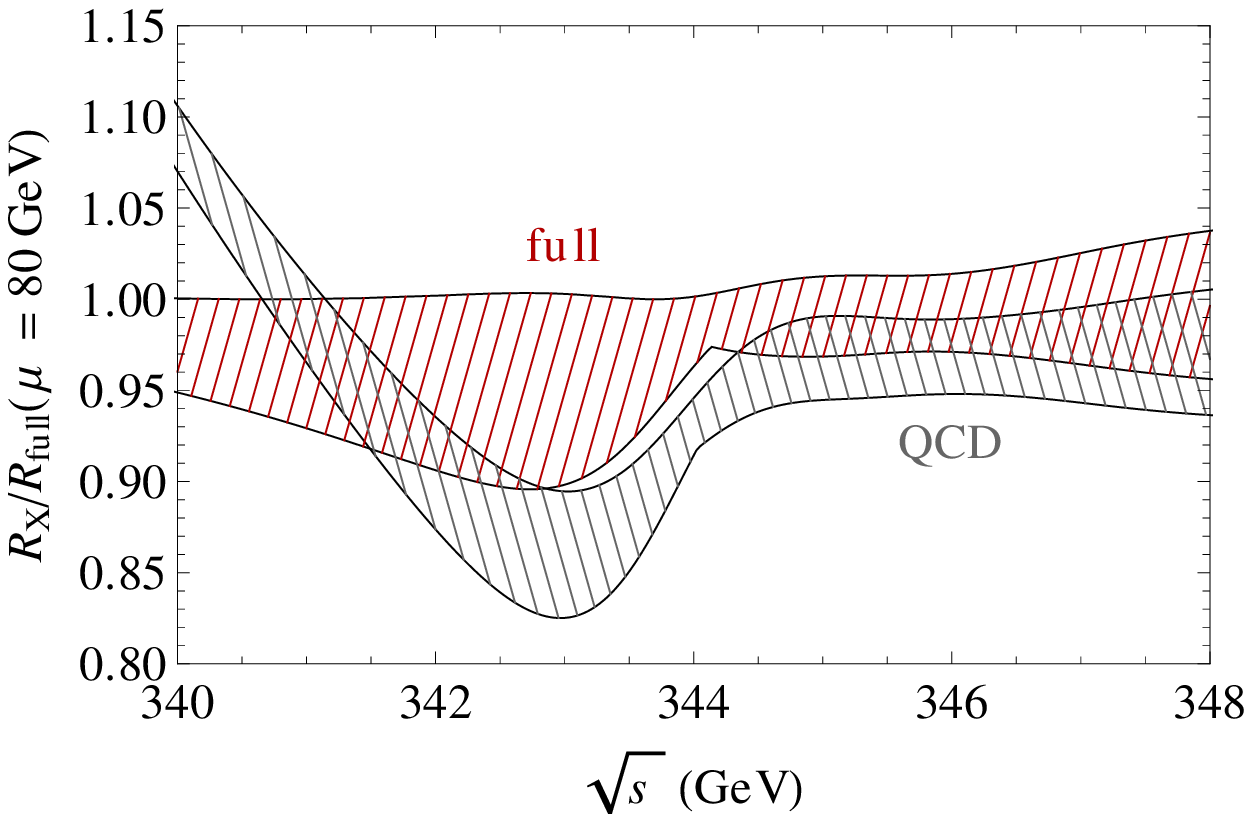}
\caption{\label{fig:scale} Overall effect of non-QCD corrections 
on the cross section. The uncertainty band is spanned by variation
of the renormalization scale $\mu\in[50\mbox{ GeV},350\mbox{ GeV}]$.
In the right plot the cross section is  normalized to the full one
at the central scale $\mu=80\mbox{ GeV}$. 
Figures from \cite{Beneke:2015lwa}.}
\vskip0.5cm
\includegraphics[width=0.48\textwidth]{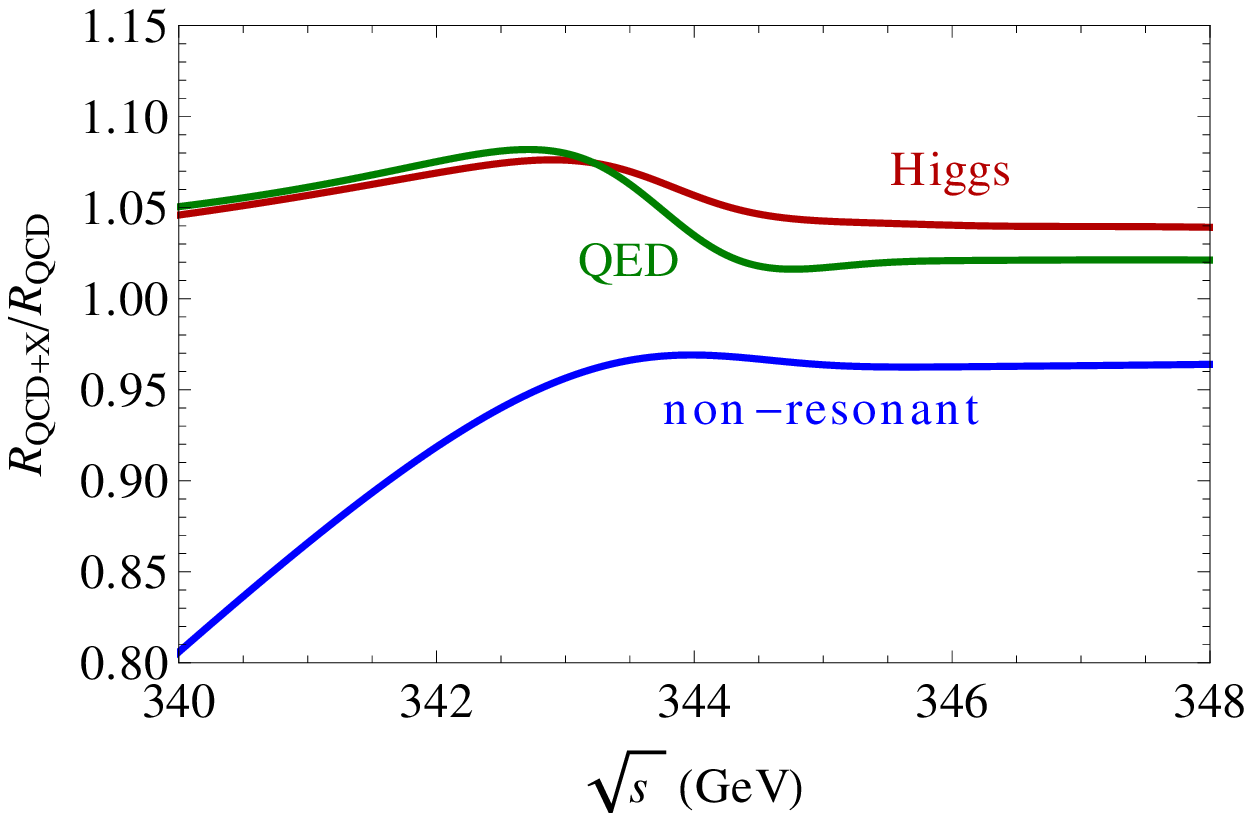}
\caption{\label{fig:corrections} Relative size of the different
non-QCD corrections to the top-pair production cross section
with respect to the pure QCD result at $\mu=80\mbox{ GeV}$. 
Figure from \cite{Beneke:2015lwa}.}
\end{figure}

\textit{QED effects.} 
The leading electroweak correction is the QED Coulomb potential 
at NLO. Its contribution can be inferred from the 
results available from the QCD calculation and has been
included in~\cite{Beneke:2015lwa}. Further electroweak effects at
NNLO~\cite{Grzadkowski:1986pm,Guth:1991ab,Hoang:2004tg,Hoang:2006pd} 
and even at NNNLO~\cite{Eiras:2006xm,Kiyo:2008mh} are known, but 
have not been included yet, since the full NNLO nonresonant correction 
is not available yet and thus no complete description of EW effects 
at this order is possible at the moment.

\begin{figure}[t]
  \centering
  \includegraphics[width=0.48\textwidth]{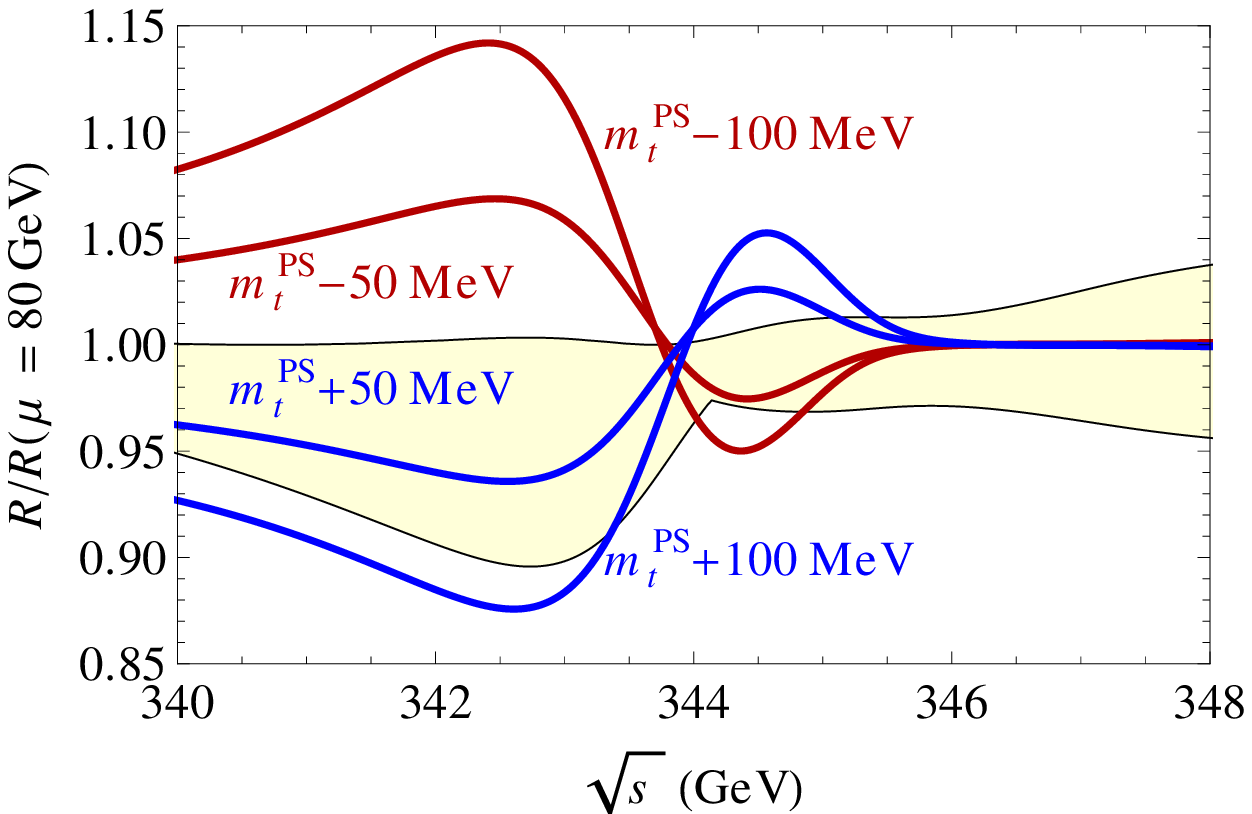}\hspace{0.2cm}
  \includegraphics[width=0.48\textwidth]{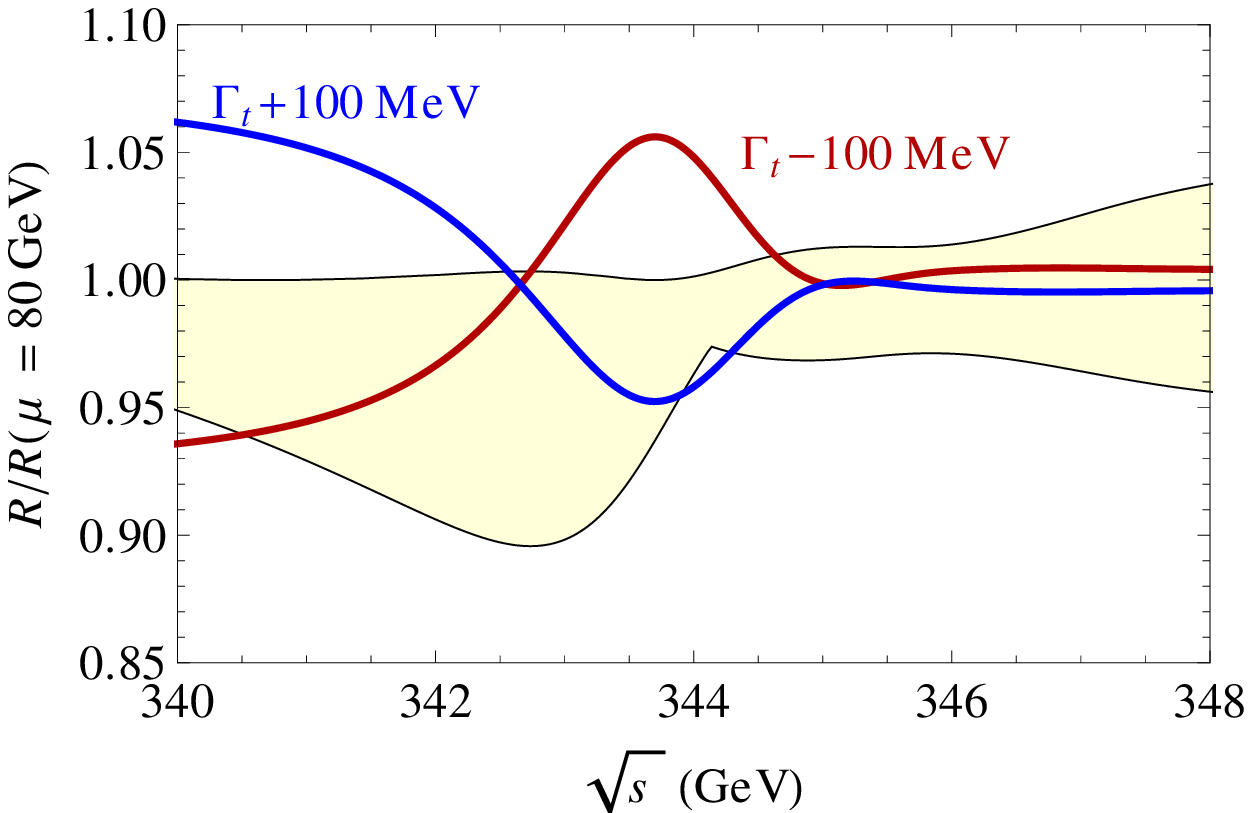}\\
  \includegraphics[width=0.48\textwidth]{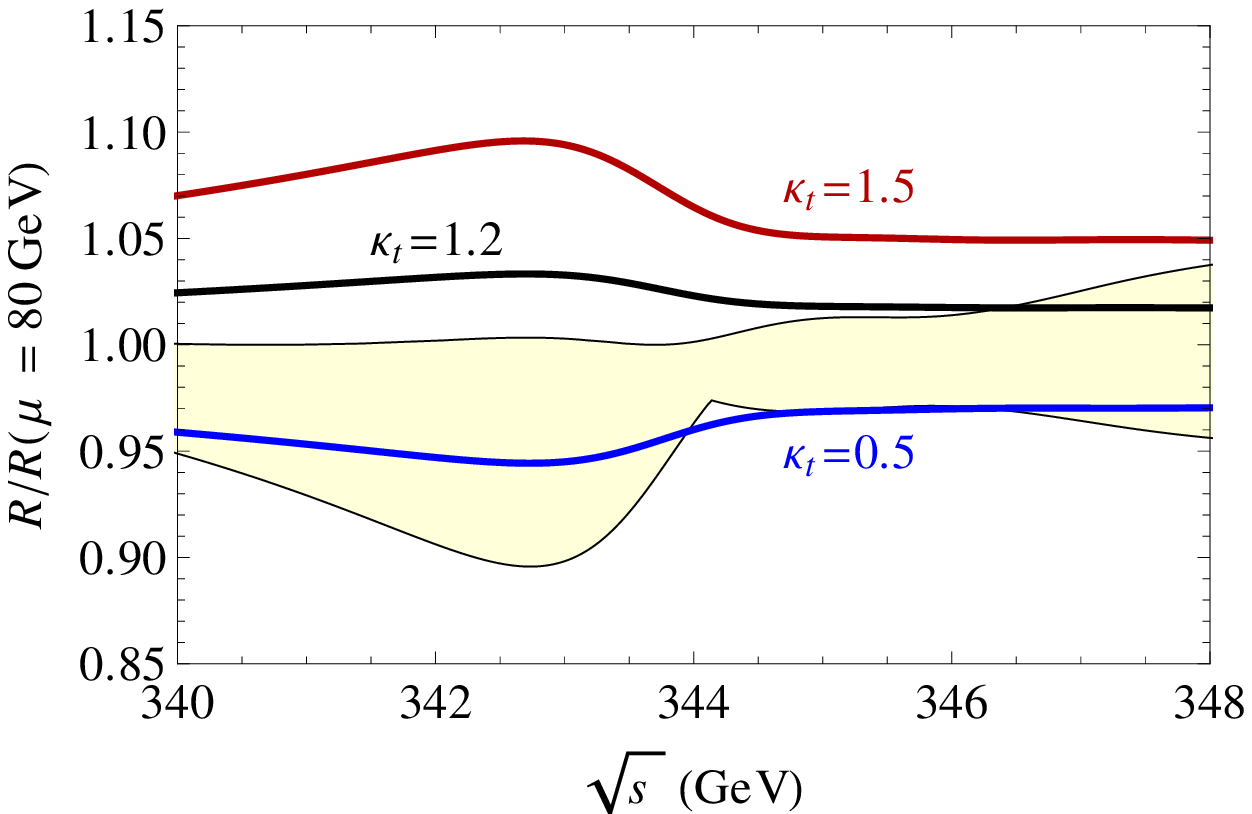}\hspace{0.2cm}
  \includegraphics[width=0.48\textwidth]{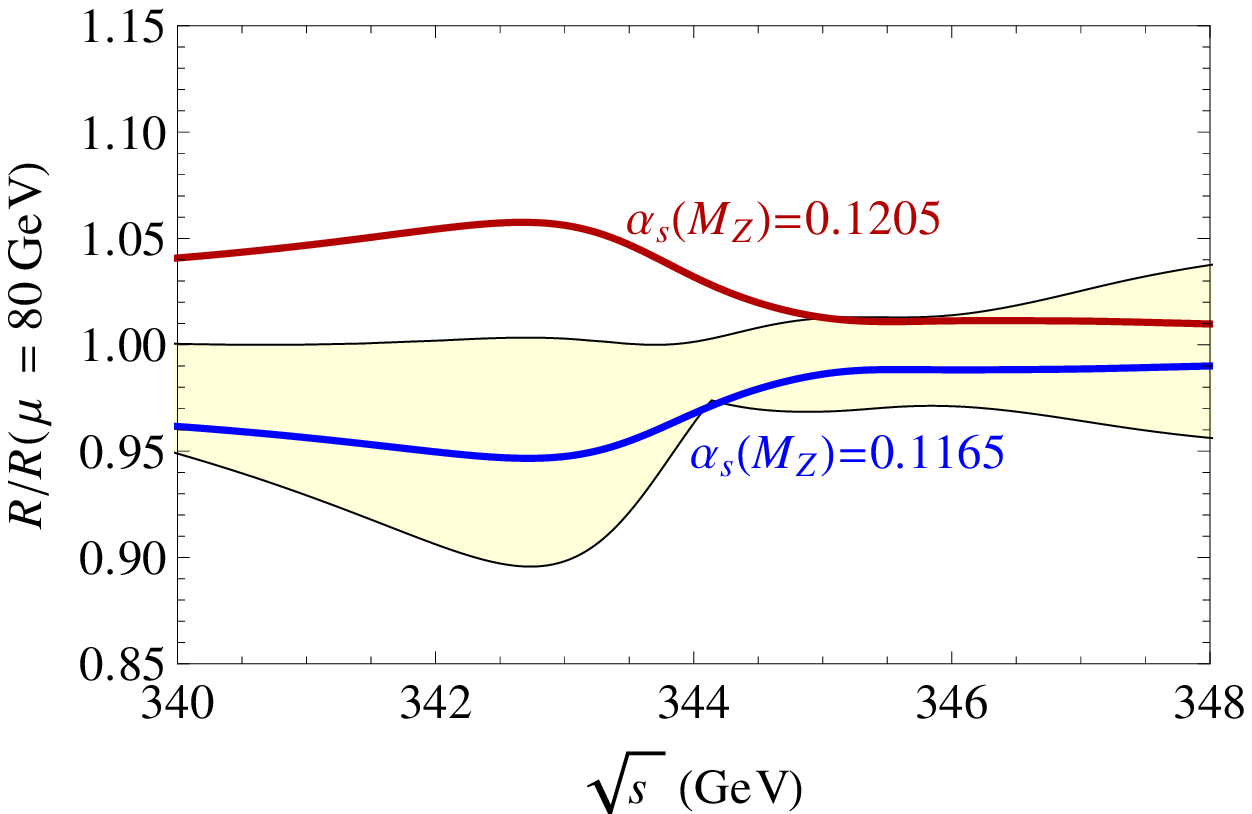}
\caption{\label{fig:parameter_dependencies} The relative change of the
cross section under variations of the top mass, width and Yukawa coupling
as well as the strong coupling constant is shown in comparison to the 
uncertainty band obtained by scale variation. Figures in the second 
row from \cite{Beneke:2015lwa}; those in the first 
row are similar to the ones shown in~\cite{Beneke:2015kwa}, except that now 
the cross section includes the P-wave and non-QCD contributions 
discussed in \cite{Beneke:2015lwa} and this proceeding.}
\end{figure}

\section{Phenomenology}
%\textit{Phenomenology.} 
\noindent We compare the non-QCD effects described above to the pure QCD 
cross section. The latter is given by the results of~\cite{Beneke:2015kwa}
to which we add the small P-wave contribution~\cite{Beneke:2013kia}. 
The net effect is shown in Figure~\ref{fig:scale}, where the uncertainty 
bands for the pure QCD and the full result are displayed 
(see \cite{Beneke:2015lwa} for the adopted parameter values). We observe
that the non-QCD contributions change the cross section by up to about 
$10\%$ and particularly affect the shape of the cross section at and 
below threshold. The shift is larger than the QCD uncertainty estimate, 
thus it is very important to include these contributions. Based on 
the shift in the peak position we estimate that the effect on the 
measurement of the top mass is approximately $50\mbox{ MeV}$, which 
is the expected total uncertainty. The separate corrections 
relative to the QCD prediction are shown in Figure~\ref{fig:corrections}.
The Higgs and QED contributions both increase the cross section by 
$4-8\%$ and $2-8\%$, respectively, since they provide an additional 
attraction between the top pair. Furthermore they shift the peak 
slightly towards smaller center-of-mass energies.\footnote{The peak 
arises from the smeared out toponium resonances, whose binding 
energy is increased by the additional attractive potentials.} 
On the other hand the nonresonant contribution is negative,  
insensitive to the special dynamics near threshold, and roughly 
energy-independent at NLO. This implies that the relative correction 
can reach up to $20\%$ below threshold, where the cross section 
becomes small.

\begin{figure}[t]
\begin{center}
\includegraphics[width=0.53\textwidth]{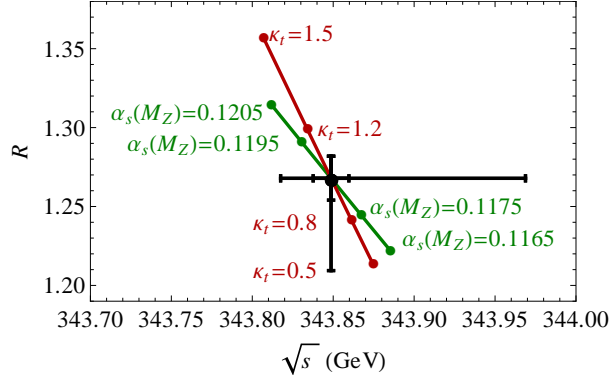}
\end{center}

\vskip-0.35cm
\caption{Changes in peak height and position due to variation of the
  Yukawa coupling (red line) and the strong coupling (green line). The
  black error bars denote the $\alpha_s$ and combined scale and
  $\alpha_s$ uncertainty for $y_t=y_t^{\mathrm{SM}}$ ($\kappa_t=1$) and
  $\alpha_s(M_Z)=0.1185$. Figure from \cite{Beneke:2015lwa}.}
\label{fig:yukaspeak}
\end{figure}

To get an idea of the physics potential of a top threshold scan 
at a future lepton collider we discuss the dependence of the cross 
section on the input parameters and compare it to the theory uncertainty. 
Relative to the full result at $\mu=80\mbox{ GeV}$ this is shown in 
Figure~\ref{fig:parameter_dependencies}. A change in the top mass 
mainly manifests itself in a horizontal shift of the cross section by twice 
that amount. An increase/decrease of the top width changes the degree 
to which the toponium resonances are smeared out and thus makes the 
peak in the cross section less/more pronounced. The parameter $\kappa_t$
parametrizes possible new physics effects in the relation between the 
top Yukawa coupling and mass $y_t=\sqrt{2}\kappa_t m_t/v$, where 
$\kappa_t=1$ corresponds to the Standard Model. Variation of $\kappa_t$
as well as the strong coupling mainly changes the normalization of the 
cross section. Due to the similar effect on the cross section the 
sensitivity to the individual parameters $\kappa_t,\alpha_s$ in a 
threshold scan is reduced if both are extracted in a simultaneous fit. 
For the peak position and height these dependences are illustrated in 
Figure~\ref{fig:yukaspeak}.  
Given the small error of the strong coupling constant it can 
also be used as an external input, in which case the added uncertainty 
relative to the scale variation is small.

%%%%%%%%%%%%%%%%%%%%%%%%%%%%%%%%%%%%%%%%%%%%%%%%%%%%%%%%%%%%%%%%%%%%%%%
%%%%%%%%%%%%%%%%%%%%%%%%%%%%%%%%%%%%%%%%%%%%%%%%%%%%%%%%%%%%%%%%%%%%%%%%%%

A rough estimate of the theory uncertainty in measurements of these 
parameters can be obtained 
by determining the parameter shifts for which the change of cross section 
lies outside of the uncertainty band in 
Figure~\ref{fig:parameter_dependencies}. This however underestimates the 
sensitivity, since in the threshold scan the cross section is measured 
for multiple points and the theory uncertainty is (at least to some
degree) correlated. A reliable estimate on the sensitivity can thus 
only be obtained by an experimental study using the full theory result.
Unfortunately this is not available yet, but existing 
studies~\cite{Martinez:2002st,Seidel:2013sqa,Horiguchi:2013wra} with 
less complete theory input suggest that the experimental uncertainties 
are about one half of the present theoretical ones for the top width and 
mass, specifically of the order of $20\mbox{ MeV}$ for the mass and  
$20-30\mbox{ MeV}$ for the width, respectively, and $0.001$ for the 
strong coupling and $5-15\%$ for 
the top Yukawa coupling. 
%From Figure~\ref{fig:parameter_dependencies} 
%and the considerations above we estimate that the theory uncertainties
%are of comparable size up to maybe a factor of two.

%

\textit{Acknowledgements.}
This work was supported by the Gottfried Wilhelm Leibniz programme
of the Deutsche For\-schungs\-gemeinschaft (DFG)
and the DFG cluster of excellence ``Origin and Structure of the Universe".


\begin{thebibliography}{99}

%%%%%%%%%%%%%%%%%%%%%%%%%%%%%%%%%%%%%%%%%%%%%%%%%%%%%%%%%%%%%%%%%%%%%
%%%%%%%%%%%%%%%%%%%%%%%%%%%% Introduction %%%%%%%%%%%%%%%%%%%%%%%%%%%
%%%%%%%%%%%%%%%%%%%%%%%%%%%%%%%%%%%%%%%%%%%%%%%%%%%%%%%%%%%%%%%%%%%%%

%\cite{ATLAS:2014wva}
\bibitem{ATLAS:2014wva}
  ATLAS and CDF and CMS and D0 Collaborations,
  %``First combination of Tevatron and LHC measurements of the top-quark mass,''
  \xref{1403.4427}{[hep-ex]}.
  %%CITATION = ARXIV:1403.4427;%%
  %246 citations counted in INSPIRE as of 19 Oct 2015

%\cite{Aad:2014kva}
\bibitem{Aad:2014kva}
  ATLAS Collaboration,
  %``Measurement of the $t\overline{t}$ production cross-section using $e\mu $ events with $b$ -tagged jets in $pp$ collisions at $\sqrt{s}=7$ and 8 TeV with the ATLAS detector,''
  \dref{10.1140/epjc/s10052-014-3109-7}{Eur.\ Phys.\ J.\ C {\bf 74} (2014) 10,  3109}
  [\xref{1406.5375}{[hep-ex]}].
  %%CITATION = ARXIV:1406.5375;%%
  %74 citations counted in INSPIRE as of 19 Oct 2015

%\cite{Chatrchyan:2013haa}
\bibitem{Chatrchyan:2013haa}
  CMS Collaboration,
  %``Determination of the top-quark pole mass and strong coupling constant from the t t-bar production cross section in pp collisions at $\sqrt{s}$ = 7 TeV,''
  \dref{10.1016/j.physletb.2013.12.009}{Phys.\ Lett.\ B {\bf 728} (2014) 496},
  \dref{10.1016/j.physletb.2014.08.040}{Corrigendum-ibid.\ B {\bf 728} (2014) 526}
  [\xref{1307.1907}{[hep-ex]}].
  %%CITATION = ARXIV:1307.1907;%%
  %82 citations counted in INSPIRE as of 19 Oct 2015

%\cite{Giudice:2015toa}
\bibitem{Giudice:2015toa}
  G.~F.~Giudice, P.~Paradisi and A.~Strumia,
  %``Indirect determinations of the top quark mass,''
  \xref{1508.05332}{[hep-ph]}.
  %%CITATION = ARXIV:1508.05332;%%
  %2 citations counted in INSPIRE as of 19 Oct 2015

%\cite{Martinez:2002st}
\bibitem{Martinez:2002st}
  M.~Martinez and R.~Miquel,
  %``Multiparameter fits to the t anti-t threshold observables at a future e+ e- linear collider,''
  \dref{10.1140/epjc/s2002-01094-1}{Eur.\ Phys.\ J.\ C {\bf 27} (2003) 49}
  [\oref{hep-ph/0207315}].
  %%CITATION = HEP-PH/0207315;%%
  %115 citations counted in INSPIRE as of 19 Aug 2015
  
  %\cite{Seidel:2013sqa}
\bibitem{Seidel:2013sqa}
  K.~Seidel, F.~Simon, M.~Tesa\v{r} and S.~Poss,
  %``Top quark mass measurements at and above threshold at CLIC,''
  \dref{10.1140/epjc/s10052-013-2530-7}{Eur.\ Phys.\ J.\ C {\bf 73} (2013) 8,  2530}\newline
  [\xref{1303.3758}{[hep-ex]}].
  %%CITATION = ARXIV:1303.3758;%%
  %40 citations counted in INSPIRE as of 19 Aug 2015
  
%\cite{Horiguchi:2013wra}
\bibitem{Horiguchi:2013wra}
  T.~Horiguchi, A.~Ishikawa, T.~Suehara, K.~Fujii, Y.~Sumino, Y.~Kiyo and H.~Yamamoto,
  %``Study of top quark pair production near threshold at the ILC,''
  \xref{1310.0563}{[hep-ex]}.
  %%CITATION = ARXIV:1310.0563;%%
  %22 citations counted in INSPIRE as of 19 Aug 2015
  
  %\cite{Beneke:2015kwa}
\bibitem{Beneke:2015kwa}
  M.~Beneke, Y.~Kiyo, P.~Marquard, A.~Penin, J.~Piclum and M.~Steinhauser,\\
  %``Next-to-next-to-next-to-leading order QCD prediction for the top anti-top S-wave pair production cross section near threshold in e+ e- annihilation,''
  \xref{1506.06864}{[hep-ph]},
  accepted for publication in Phys. Rev. Lett.
  %%CITATION = ARXIV:1506.06864;%%
  %1 citations counted in INSPIRE as of 18 Aug 2015

  %\cite{Hoang:2000yr}
\bibitem{Hoang:2000yr}
  A.~H.~Hoang {\it et al.},
  %``Top - anti-top pair production close to threshold: Synopsis of recent NNLO results,''
  \dref{10.1007/s1010500c0003}{Eur.\ Phys.\ J.\ direct C {\bf 3} (2000) 1}
  [\oref{hep-ph/0001286}].
  %%CITATION = HEP-PH/0001286;%%
  %296 citations counted in INSPIRE as of 18 Aug 2015

%\cite{Beneke:2015lwa}
\bibitem{Beneke:2015lwa}
  M.~Beneke, A.~Maier, J.~Piclum and T.~Rauh,
  %``Higgs effects in top anti-top production near threshold in $e^+e^−$ annihilation,''
  \dref{10.1016/j.nuclphysb.2015.07.034}{Nucl.\ Phys.\ B {\bf 899} (2015) 180}\newline
  [\xref{1506.06865}{[hep-ph]}].
  %%CITATION = ARXIV:1506.06865;%%
  %1 citations counted in INSPIRE as of 15 Oct 2015
  
%%%%%%%%%%%%%%%%%%%%%%%%%%%%%%%%%%%%%%%%%%%%%%%%%%%%%%%%%%%%%%%%%%%%%
%%%%%%%%%%%%%%%%%%%%%%%%%%%%%% Dynamics %%%%%%%%%%%%%%%%%%%%%%%%%%%%%
%%%%%%%%%%%%%%%%%%%%%%%%%%%%%%%%%%%%%%%%%%%%%%%%%%%%%%%%%%%%%%%%%%%%%
  
%\cite{Beneke:2013jia}
\bibitem{Beneke:2013jia}
  M.~Beneke, Y.~Kiyo and K.~Schuller,
  %``Third-order correction to top-quark pair production near threshold I. Effective theory set-up and matching coefficients,''
  \xref{1312.4791}{[hep-ph]}.
  %%CITATION = ARXIV:1312.4791;%%
  %20 citations counted in INSPIRE as of 18 août 2015
  
%%%%%%%%%%%%%%%%%%%%%%%%%%%%%%%%%%%%%%%%%%%%%%%%%%%%%%%%%%%%%%%%%%%%%
%%%%%%%%%%%%%%%%%%%%%%%%%%%%%% P wave %%%%%%%%%%%%%%%%%%%%%%%%%%%%%%%
%%%%%%%%%%%%%%%%%%%%%%%%%%%%%%%%%%%%%%%%%%%%%%%%%%%%%%%%%%%%%%%%%%%%%
  

%\cite{Beneke:2013kia}
\bibitem{Beneke:2013kia}
  M.~Beneke, J.~Piclum and T.~Rauh,
  %``P-wave contribution to third-order top-quark pair production near threshold,''
  \dref{10.1016/j.nuclphysb.2014.01.015}{Nucl.\ Phys.\ B {\bf 880} (2014) 414}
  [\xref{1312.4792}{[hep-ph]}].
  %%CITATION = ARXIV:1312.4792;%%
  %2 citations counted in INSPIRE as of 24 Apr 2015

%%%%%%%%%%%%%%%%%%%%%%%%%%%%%%%%%%%%%%%%%%%%%%%%%%%%%%%%%%%%%%%%%%%%%
%%%%%%%%%%%%%%%%%%%%%%%%%%%%%%% Higgs %%%%%%%%%%%%%%%%%%%%%%%%%%%%%%%
%%%%%%%%%%%%%%%%%%%%%%%%%%%%%%%%%%%%%%%%%%%%%%%%%%%%%%%%%%%%%%%%%%%%%

\bibitem{Strassler:1990nw}
M.~J. Strassler and M.~E. Peskin, 
%{\it The heavy top quark threshold: {QCD} and the {Higgs}},  
\dref{10.1103/PhysRevD.43.1500}{Phys. Rev. {\bf D43} (1991) 1500}.
%%CITATION = PHRVA,D43,1500;%%

\bibitem{Harlander:1995dp}
  R.~Harlander, M.~Je\.zabek and J.~H.~K\"uhn,
%  {\it Higgs effects in top quark pair production,}
  \href{http://www.actaphys.uj.edu.pl/_cur/store/vol27/pdf/v27p1781.pdf}{Acta Phys.\ Polon.\  {\bf B27} (1996) 1781}
  [\oref{hep-ph/9506292}].

  %\cite{Grzadkowski:1986pm}
\bibitem{Grzadkowski:1986pm}
  B.~Grzadkowski, J.~H.~K\"uhn, P.~Krawczyk and R.~G.~Stuart,
  %``Electroweak Corrections on the Toponium Resonance,''
  \dref{10.1016/0550-3213(87)90245-8}{Nucl.\ Phys.\ B {\bf 281} (1987) 18}.
  %%CITATION = NUPHA,B281,18;%%
  %62 citations counted in INSPIRE as of 24 Apr 2015

  %\cite{Guth:1991ab}
\bibitem{Guth:1991ab}
  R.~J.~Guth and J.~H.~K\"uhn,
  %``Top quark threshold and radiative corrections,''
  \dref{10.1016/0550-3213(92)90196-I}{Nucl.\ Phys.\ B {\bf 368} (1992) 38}.
  %%CITATION = NUPHA,B368,38;%%
  %52 citations counted in INSPIRE as of 24 Apr 2015

 %\cite{Eiras:2006xm}
\bibitem{Eiras:2006xm}
  D.~Eiras and M.~Steinhauser,
  %``Complete Higgs mass dependence of top quark pair threshold production to order alpha alpha(s),''
  \dref{10.1016/j.nuclphysb.2006.09.010}{Nucl.\ Phys.\ B {\bf 757} (2006) 197}
  [\oref{hep-ph/0605227}].
  %%CITATION = HEP-PH/0605227;%%
  %12 citations counted in INSPIRE as of 09 Dec 2013
  
%%%%%%%%%%%%%%%%%%%%%%%%%%%%%%%%%%%%%%%%%%%%%%%%%%%%%%%%%%%%%%%%%%%%%
%%%%%%%%%%%%%%%%%%%%%%%%%%%%%%% Nonres %%%%%%%%%%%%%%%%%%%%%%%%%%%%%%
%%%%%%%%%%%%%%%%%%%%%%%%%%%%%%%%%%%%%%%%%%%%%%%%%%%%%%%%%%%%%%%%%%%%%

%\cite{Fadin:1987wz}
\bibitem{Fadin:1987wz}
  V.~S.~Fadin and V.~A.~Khoze,
  %``Threshold Behavior of Heavy Top Production in e+ e- Collisions,''
  \href{http://www.jetpletters.ac.ru/ps/1234/article_18631.shtml}{JETP Lett.\  {\bf 46} (1987) 525}
   [\href{http://www.jetpletters.ac.ru/ps/167/article_2833.shtml}{Pisma Zh.\ Eksp.\ Teor.\ Fiz.\  {\bf 46} (1987) 417}].
  %%CITATION = JTPLA,46,525;%%
  %266 citations counted in INSPIRE as of 19 Aug 2015

%\cite{Beneke:2003xh}
\bibitem{Beneke:2003xh}
  M.~Beneke, A.~P.~Chapovsky, A.~Signer and G.~Zanderighi,
  %``Effective theory approach to unstable particle production,''
  \dref{10.1103/PhysRevLett.93.011602}{Phys.\ Rev.\ Lett.\  {\bf 93} (2004) 011602}
  [\oref{hep-ph/0312331}].
  %%CITATION = HEP-PH/0312331;%%
  %78 citations counted in INSPIRE as of 19 août 2015
  
  %\cite{Beneke:2004km}
\bibitem{Beneke:2004km}
  M.~Beneke, A.~P.~Chapovsky, A.~Signer and G.~Zanderighi,
  %``Effective theory calculation of resonant high-energy scattering,''
  \dref{10.1016/j.nuclphysb.2004.03.016}{Nucl.\ Phys.\ B {\bf 686} (2004) 205}
  [\oref{hep-ph/0401002}].
  %%CITATION = HEP-PH/0401002;%%
  %68 citations counted in INSPIRE as of 19 août 2015
  
  %\cite{Beneke:2010mp}
\bibitem{Beneke:2010mp}
  M.~Beneke, B.~Jantzen and P.~Ruiz-Femen\'ia,
  %``Electroweak non-resonant NLO corrections to e+ e- -> W+ W- b bbar in the t tbar resonance region,''
  \dref{10.1016/j.nuclphysb.2010.07.006}{Nucl.\ Phys.\ B {\bf 840} (2010) 186}\newline
  [\xref{1004.2188}{[hep-ph]}].
  %%CITATION = ARXIV:1004.2188;%%
  %21 citations counted in INSPIRE as of 24 Apr 2015
  
  %\cite{Penin:2011gg}
\bibitem{Penin:2011gg}
  A.~A.~Penin and J.~H.~Piclum,
  %``Threshold production of unstable top,''
  \dref{10.1007/JHEP01(2012)034}{JHEP {\bf 1201} (2012) 034}
  [\xref{1110.1970}{[hep-ph]}].
  %%CITATION = ARXIV:1110.1970;%%
  %17 citations counted in INSPIRE as of 24 Apr 2015
  
  %\cite{Jantzen:2013gpa}
\bibitem{Jantzen:2013gpa}
  B.~Jantzen and P.~Ruiz-Femen\'ia,
  %``Next-to-next-to-leading order nonresonant corrections to threshold top-pair production from $e^+e^-$ collisions: Endpoint-singular terms,''
  \dref{10.1103/PhysRevD.88.054011}{Phys.\ Rev.\ D {\bf 88} (2013) 5,  054011}
  [\xref{1307.4337}{[hep-ph]}].
  %%CITATION = ARXIV:1307.4337;%%
  %7 citations counted in INSPIRE as of 24 Apr 2015
  
  %\cite{Ruiz-Femenia:2014ava}
\bibitem{Ruiz-Femenia:2014ava}
  P.~Ruiz-Femen\'ia,
  %``First estimate of the NNLO nonresonant corrections to top-antitop threshold production at lepton colliders,''
  \dref{10.1103/PhysRevD.89.097501}{Phys.\ Rev.\ D {\bf 89} (2014) 9,  097501}
  [\xref{1402.1123}{[hep-ph]}].
  %%CITATION = ARXIV:1402.1123;%%
  %1 citations counted in INSPIRE as of 24 Apr 2015
  
  
%%%%%%%%%%%%%%%%%%%%%%%%%%%%%%%%%%%%%%%%%%%%%%%%%%%%%%%%%%%%%%%%%%%%%
%%%%%%%%%%%%%%%%%%%%%%%%%%%%%%% QED %%%%%%%%%%%%%%%%%%%%%%%%%%%%%%%%%
%%%%%%%%%%%%%%%%%%%%%%%%%%%%%%%%%%%%%%%%%%%%%%%%%%%%%%%%%%%%%%%%%%%%%

  %\cite{Hoang:2004tg}
\bibitem{Hoang:2004tg}
  A.~H.~Hoang and C.~J.~Rei\ss er,
  %``Electroweak absorptive parts in NRQCD matching conditions,''
  \dref{10.1103/PhysRevD.71.074022}{Phys.\ Rev.\ D {\bf 71} (2005) 074022}
  [\oref{hep-ph/0412258}].
  %%CITATION = HEP-PH/0412258;%%
  %49 citations counted in INSPIRE as of 24 Apr 2015
  
 %\cite{Hoang:2006pd}
\bibitem{Hoang:2006pd}
  A.~H.~Hoang and C.~J.~Rei\ss er,
  %``On electroweak matching conditions for top pair production at threshold,''
  \dref{10.1103/PhysRevD.74.034002}{Phys.\ Rev.\ D {\bf 74} (2006) 034002}
  [\oref{hep-ph/0604104}].
  %%CITATION = HEP-PH/0604104;%%
  %24 citations counted in INSPIRE as of 24 Apr 2015
    
  %\cite{Kiyo:2008mh}
\bibitem{Kiyo:2008mh}
  Y.~Kiyo, D.~Seidel and M.~Steinhauser,
  %``O(alpha alpha(s)) corrections to the gamma t anti-t vertex at the top quark threshold,''
  \dref{10.1088/1126-6708/2009/01/038}{JHEP {\bf 0901} (2009) 038}
  [\xref{0810.1597}{[hep-ph]}].
  %%CITATION = ARXIV:0810.1597;%%
  %13 citations counted in INSPIRE as of 24 Apr 2015

\end{thebibliography}
\end{document}